\documentclass[%
reprint,
 amsmath,amssymb,
 apl,
]{revtex4-1}

\usepackage{graphicx}
\usepackage{dcolumn}
\usepackage{bm}

\usepackage{xcolor}

\newcommand{\red}{\textcolor{black}}

\begin{document}

\title{Terahertz pulsed photogenerated current in microdiodes at room temperature}

\author{Marjan Ilkov}
 \affiliation{School of Science and Engineering, Reykjavik University, 
              Menntavegur 1, IS-101 Reykjavik, Iceland}

\author{Kristinn Torfason}
 \affiliation{School of Science and Engineering, Reykjavik University, 
              Menntavegur 1, IS-101 Reykjavik, Iceland}
 
\author{Andrei Manolescu}
 \affiliation{School of Science and Engineering, Reykjavik University, 
              Menntavegur 1, IS-101 Reykjavik, Iceland}

\author{\' Ag\' ust~Valfells}
 \affiliation{School of Science and Engineering, Reykjavik University, 
              Menntavegur 1, IS-101 Reykjavik, Iceland}


\begin{abstract} 
Space-charge modulation of the current in a vacuum diode under photoemission leads to the formation of beamlets with time periodicity corresponding to THz frequencies.  We investigate the effect of the emitter temperature and internal space-charge forces on the formation and persistence of the beamlets. \red{We find that temperature effects are most important for beam degradation at low values of the applied electric field, whereas at higher fields intra-beamlet  space-charge forces are dominant.} The current modulation is most robust when there is only one beamlet present in the diode gap at a time, corresponding to a macroscopic version of the Coulomb blockade.  It is shown that a vacuum microdiode can operate quite well as a tunable THz oscillator at room temperature with an applied electric field above 10 MV/m and a diode gap of the order of 100 nanometers.

\end{abstract}

\pacs{
52.59.Rz, 
73.23.Hk, 
29.20.    
}

\maketitle


\textit{Introduction.} Space-charge limited flow in a diode has been a subject of
investigation for over a century \cite{child1911,langmuir1913,
birdsall1966}. Nonetheless, this field of research remains quite
fertile in terms of interesting results. The past fifteen years have
seen considerable work on extending the classical Child-Langmuir
law \cite{luginsland2002}, e.g. for very small structures
\cite{ang2006, ang2003, zhu2011}, finite emitter area and pulse length
\cite{valfells2002,lau2001,ang2007}, space-charge limited field emission
\cite{feng2006,rokhlenko2010,torfason2015}, as well as development of
novel scaling laws and investigations into the nature of space-charge
limited flow \cite{akimov2001, caflisch2012,griswold2010,zhu2013}.

\begin{figure}
\begin{center}
\vspace{-8 mm}
\includegraphics [scale=0.34] {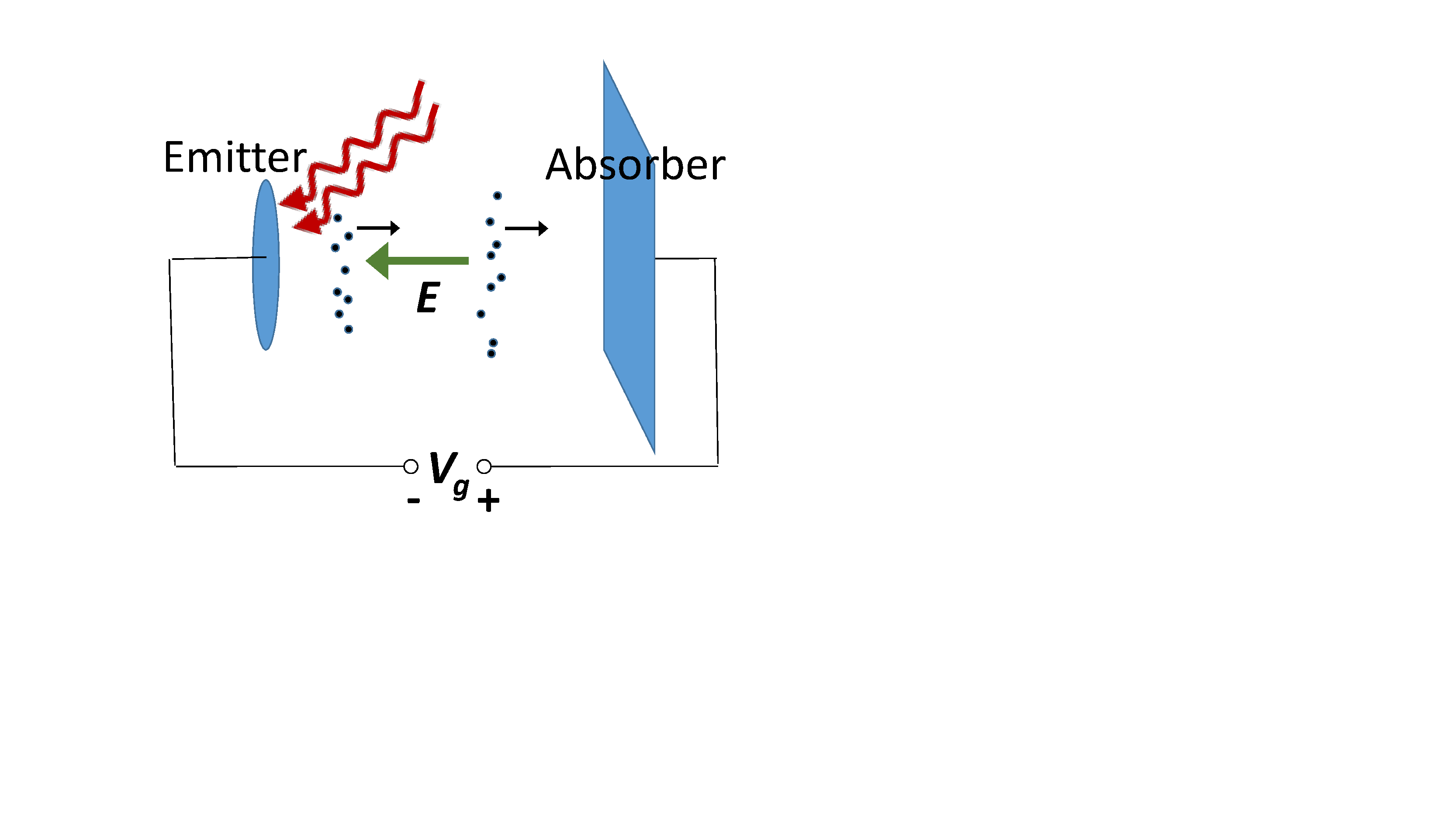}
\end{center}
\vspace{-30mm}
\caption{The device is a vacuum diode with a disk-shaped emitter on an infinite cathode under a strong laser pulse. The electrons are extracted in bunches and are driven to the absorber (anode) by the electric field $E$.
}
\vspace{-6mm}
\label{diode}
\end{figure}

Recent research has been on the formation of regular
electron bunches under space-charge limited conditions in microdiodes
\cite{pedersen2011}. Simulations and analysis indicate that in diodes
with gap spacing from hundreds of nm to a couple of microns, and applied
potential of the order of Volts, space-charge effects can lead to the
formation of well defined electron bunches at the cathode, that lead
to a regularly varying anode current, Fig. \ref{diode}. The frequency of this current
is in the THz band and can be tuned by simply varying the applied
electric field inside the diode \cite{jonsson2013}. Furthermore,
it appears that higher power can be extracted from the diode if the
cathode is covered with an array of nanoscale emitting „dots“ each
emitting a stream of beamlets that is synchronized with the beamlets
from neighboring emitters \cite{ilkov2015}.  \red{A recent paper that
presents a study on the maximal charge allowed for stable injection of
electron bunches into a diode with a regular interval \cite{liu2015} has
some interesting parallels to our previous work, although the geometry
and operating parameters are not directly comparable}.

Although this is an interesting idea for THz generation and applications
\cite{siegel2002a,booske2011,jepsen2011, tonouchi2007}, there is still
a fundamental issue that needs to be addressed.  Under what circumstances
can one expect to still see the time dependent structure intact? Two particular
candidate causes for degradation of the beamlets are: velocity spread (or
emittance); and Coulomb forces causing beamlets to expand and merge. In
this paper we will investigate the issue by looking in detail at how
diode gap spacing, applied potential and velocity spread of electrons
emitted from the cathode influence pulse formation and degradation.

\textit{Physical and simulation setup.} The system under study is an infinite 
parallel plate vacuum diode, with gap spacing $D$, sketched in Fig.~\ref{diode}. 
Emission from the cathode is restricted to a circular area of radius R = 150 nm. 
A voltage is applied across the diode gap, $V_g$, and it is kept
constant throughout one simulation. The vacuum field for a diode without
any electrons in the gap is $E$. 

Emission is not source limited, so we assume unlimited supply
of electrons. This creates in turn, space-charge limited current in
the diode gap. The flow of current is as follows: Once the device is
turned on, a bunch starts to form in front of the emitter. To extract an
electron from a particular emitter location, the surface field must point
towards the emitter, i.e.\ that spot must be favorable for emission.
If so, emission is permitted and the emitted electron can probably be
accelerated towards the absorber. We say probably because,
 the electrons already in the gap may position themselves in such
a way that, later on, they can push this electron back into the emitter.
After this, a new, random spot is checked on the cathode. If the new spot
is favorable, as before, an electron is placed 1 nm above the surface
of the cathode. This is done to ensure that the normal component of
the electric field on the cathode due to the newly placed electron is
nonzero. In reality such emitters will have surface irregularities which
will be larger than 1 nm which permits us to choose such distance for
electron placement. \red{The numerical results are not sensitive 
to the exact height of placement as long as it is much smaller than the gap spacing 
\cite{jonsson2013}.} If an unfavorable spot is found, a new one is randomly
picked and tested.  If in hundred consecutive checks, no favorable spot
is found, the cathode is deemed unfavorable for emission and the code
continues to the next step which is the advancement of the electrons.

Now, the forces  between all electrons in the diode gap are calculated
together with the forces due to the applied vacuum field. These forces, coupled
with the basic Verlet method \cite{grubmuller1991, hairer2003,marx2009},
advance the electrons for the duration of one time step, $t_s = 2.5\times
10^{-16}$ s. After that, it is checked whether electrons have entered
the anode, or if due to Coulomb forces they are pushed back into the
cathode by the electrons ahead. If they did, the time and place of
absorption is noted and the electrons are taken out of the system. In
this moment one time step is finished, and a new time step begins with
the emission process all over again.  The complete explanation and
setup of the system is given in detail in previous work, but only for initial
velocities of electrons equal to zero \cite{jonsson2013,pedersen2011}.
In this present paper we consider non-zero initial velocities.

For the photoemission process \cite{jensen2009photoemission,jensen2007photoemission} 
we will divide the emission region into two
separate regions: metal and vacuum. The energy $\varepsilon$ of electrons inside the
metal corresponds to the free electron gas and it is governed by
the Fermi-Dirac statistics.  The energy of the vacuum state is 
$E_{\rm F} + \phi_{\rm eff}$, where $E_{\rm F}$ is the Fermi energy and
\begin{equation}
\phi_{\rm eff} = \phi_{\rm w} - \phi_{\rm S} 
= \phi_{\rm w}-e\sqrt{\frac{eE}{4\pi\epsilon_0}} \ .
\end{equation}
$\phi_{\rm w}$ is the work function, $\phi_{\rm S}$ is the Schottky work
function, and $\phi_{\rm eff}$ is the height of the photoemission potential barrier.  The
energy after emission is equal to $\varepsilon + \hbar\omega - \phi_{\rm
eff}$. If the laser used for photoemission can be tuned
such that $\hbar\omega =~\phi_{\rm w}$, then the electron
energy will be distributed by the Maxwell-Boltzmann statistics
\cite{dowell2009,neppl2012,jensen2010emittance,jensen2012space,jensen2014emittance}.
The fact that the Fermi-Dirac distribution is virtually identical to
the Maxwell-Boltzmann distribution for energies higher than $1.005E_{\rm
F}$  at room temperature justifies using Maxwell-Boltzmann for the
electron density of states \cite{jensen2010emittance}. \red{A
mismatch between the laser energy and the work function is allowable as long
as it is much smaller that the thermal energy $k_B T$, $k_B$ being
the Boltzmann constant and $T$ the temperature.} Also, the electrons
available for emission come only from the highest energy states. The aim
of this paper is to try to model a room temperature device, therefore the
interest is in the high temperature tail of the distributions.
We consider the Maxwell-Boltzmann distribution of electron velocities $v_i$ 
at the emitter surface, in the spatial direction $i=\{x,y,z\}$,
\begin{equation}
f_v(v_i) = \sqrt{\frac{m}{2\pi k_{\rm B} T}}\exp\left(\frac{-mv_i^2}{2k_BT}\right) \ ,
\label{MB}
\end{equation}
$m$ being the mass of the electron. 
The velocities in the propagation direction $z$,
i.\ e. from left to right in Fig. \ref{diode}, can only have positive
values, whereas those in the perpendicular directions $x$ and $y$
can take any value.

\textit{Results and analysis.} We will now show the results in parameter space. 
The temperatures tested are $T=0,1,2,4,8,16,32,64,128,256,300$ K.
They are shown on the horizontal axes in the following figures.
The gap spacing, i.e.\ the distance between emitter and absorber,
has values $D = 50,100,200,400$ nm, shown on the vertical axes. 
Each figure is made with a fixed value of the electric field, 
$E=1,2,4,8,16,32,64,128$ MV/m. 

For each combination of parameters we run six simulations. The total simulated time 
for each case was $t_{\rm tot}=3.75\times10^{-10}$ s. 
For the systems with lowest frequencies ($f_{\rm min} \approx 0.8$ THz) we get $\sim$ 300 bunching events 
and for the highest frequencies ($f_{\rm max} \approx 2.2$ THz) about 800 bunching events in total.
%
\begin{figure}[h]
\begin{center}
\vspace{-3mm}
\includegraphics [width=8.8 cm] {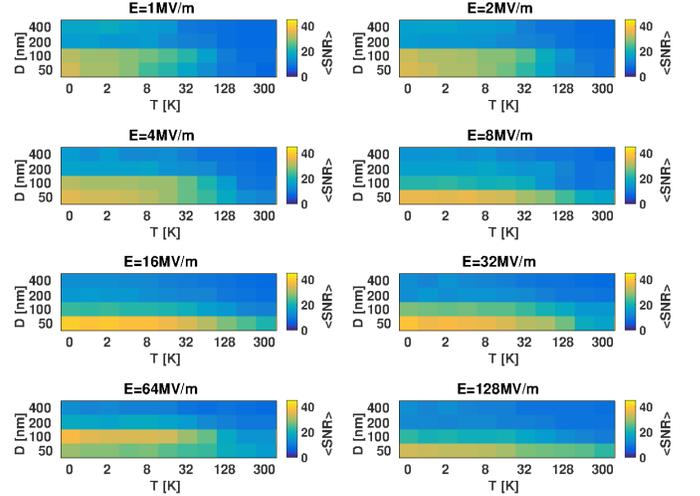}
\end{center}
\vspace{-6mm}
\caption{The averaged signal-to-noise ratio $<\rm SNR>$.}
\label{avg_snr}
\end{figure}

\textit{Averaged signal-to-noise ratio (SNR).} \red{The averaged
SNR is a good measure of the quality of the beam
modulation.  It is calculated using the peak amplitude of the Fourier
spectrum of the diode current $A_s$ and the average amplitude of the
noise $A_n$, as  ${{\rm SNR [dB]} = 20\log_{10}(A_s/A_n)}$. The computational 
algorithm uses unfiltered Fourier transform} and searches for the
highest peak in a predetermined region. This region can be found
through the parameterization $f{\rm [Hz]}=A \times \left(E{\rm
[V/m]}\right)^{\alpha}$
given in \cite{jonsson2013} where $f$ is the frequency measured in Hz
(with the applied vacuum field in units of V/m). The parameters $A$
and $\alpha$ depend on the size of the emitter, i.e. the radius of
the disk.  Table \ref{table1} shows values of these parameters for three 
emitter radii.
\begin{table} [h]
\centering
\vspace{-5 mm}
\caption{Magnitude of the parameters $A$ and $\alpha$ 
for different values of the emitter radius.}
\renewcommand{\arraystretch}{1.2}
\begin{tabular}{|l|l|l|ll}
\cline{1-3}
Emitter radius [nm]     & A                 & $\alpha$ &  &  \\ \cline{1-3}
50                      & $779 \times 10^6$ & 0.539    &  &  \\ \cline{1-3}
100                     & $326 \times 10^6$ & 0.580    &  &  \\ \cline{1-3}
250                     & $257 \times 10^6$ & 0.575    &  &  \\ \cline{1-3}
\end{tabular}
\label{table1}
\vspace{-3 mm}
\end{table}   
If the spectrum contains higher harmonics, as it often does, the
algorithm chooses only the first harmonic for the analysis. Although
higher harmonics can carry a significant portion of the energy, keeping
with the conservative nature of this research, only the first, and always
highest, peak is chosen.

\red{From Fig.\ \ref{avg_snr} it can be seen that the transition from
the parameter space where the SNR is high to that where it is low
is more dependent on temperature (an almost vertical boundary) for
low values of the applied field and becomes more dependant on the gap
spacing (an almost horizontal boundary) as field strength increases. The
reason for this is that the number of electrons per bunch increases with
applied electric field, which leads to greater intra-beam space-charge
forces, which in turn come to dominate thermal effects as the cause
of degradation.  Although the definition of what constitutes a good
signal-to-noise ratio is somewhat arbitrary and application dependant,
we consider the SNR to be good if it is above 20 dB.}

\red{We notice that the highest SNR for $E=64$ MV/m occurs at $D=100$ nm
instead of 50 nm as for other fields. This is because the degradation of
the beamlets is a complex process. A long transit time allows beamlets to
spread, but so does a greater number of electrons per bunch. However, 
since such an effect is not apparent at room temperature a detailed
investigation is beyond the purpose of the present work.  }

\begin{figure} [h]
\begin{center}
\vspace{-3mm}
\includegraphics [width=8.8 cm] {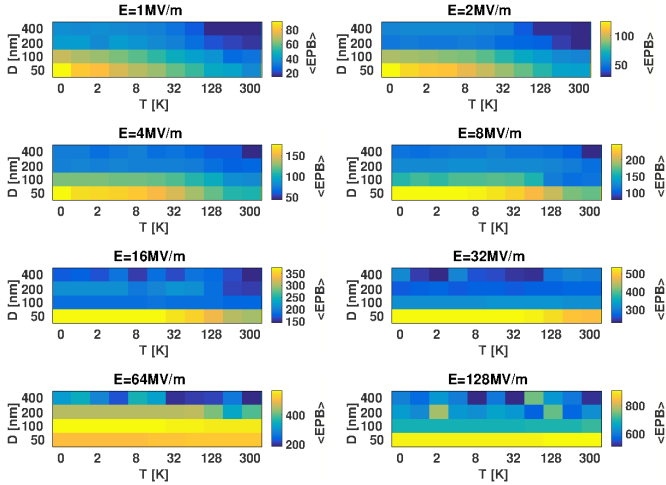}
\end{center}
\vspace{-6mm}
\caption{Average number of electrons per bunch $<{\rm EPB}>$.  }
\label{avg_epb}
\vspace{-2mm}
\end{figure}

\begin{figure}[h]
\begin{center}
\vspace{-3mm}
\includegraphics [width=8.8 cm] {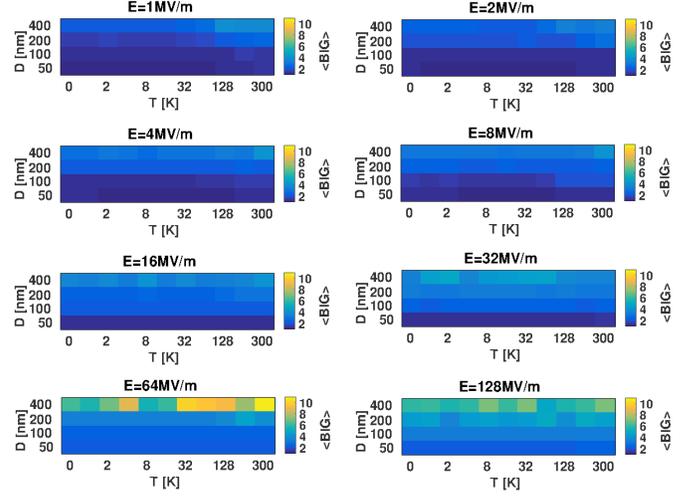}
\end{center}
\vspace{-6mm}
\caption{Average number of bunches in the gap $<{\rm BIG}>$.}
\label{avg_big}
\vspace{-2mm}
\end{figure}

\textit{Average number of electrons per bunch (EPB).} 
The number of EPB increases with increasing electric
field due to the fact that more electrons per bunch are needed to create
field reversal at the emitter surface \cite{pedersen2011}.
\red{The total number of electrons in the gap is proportional to the applied
field (see Eq.\ (4)  of \cite{valfells2002} and Eq.\ (7) of \cite{jonsson2013}). 
For a constant field the total number of electrons is nearly constant, but a larger
gap can accomodate more bunches, leading to fewer EPB.}
The number of EPB also increases 
with decreasing the temperature, since the bunches have more definitive 
structure and less smearing, Fig.\ \ref{avg_epb}. This
creates robust and solid bunches which block the cathode for a long
time before another bunch is formed.  As the temperature is increased
bunch destruction begins much faster after formation. Because no real
blockade on the cathode is created another bunch forms much sooner than it
would happen at lower temperatures.  This might mean more bunches in the
gap with increasing temperature, as can be seen in Fig.\ \ref{avg_big},
but each bunch has fewer electrons and the current begins to resemble a continuous stream.


\textit{Average number of bunches in the gap (BIG).} The number of BIG
increases with increasing vacuum field, gap spacing, and more subtly with temperature.  As the vacuum field
increases, the total number of electrons in the gap also grows and this
means more bunches in the gap. This comes from the simple Child-Langmuir
law \cite{langmuir1913}. Similarly, increasing gap spacing means more
space and more bunches can be accommodated into it.  As the temperature
increases, the initial velocity of the electrons increases as well. This lowers
the blocking potential from the charge already present in the gap, the 
emission is facilitated,
and the virtual cathode from the electrons already present in
the gap can be more easily overcome. If the temperature is low,
all electrons exit the emitter with similar velocities and the bunch is
more narrow in the direction of propagation. 
\red {The $<{\rm BIG}>$ (average
number per bunch) shows that the signal quality decreases with 
increasing this number. Which means that THz signal
is the strongest at $<{\rm BIG}>\approx1$.}

\begin{figure} [h]
\begin{center}
\includegraphics [width=8.8 cm] {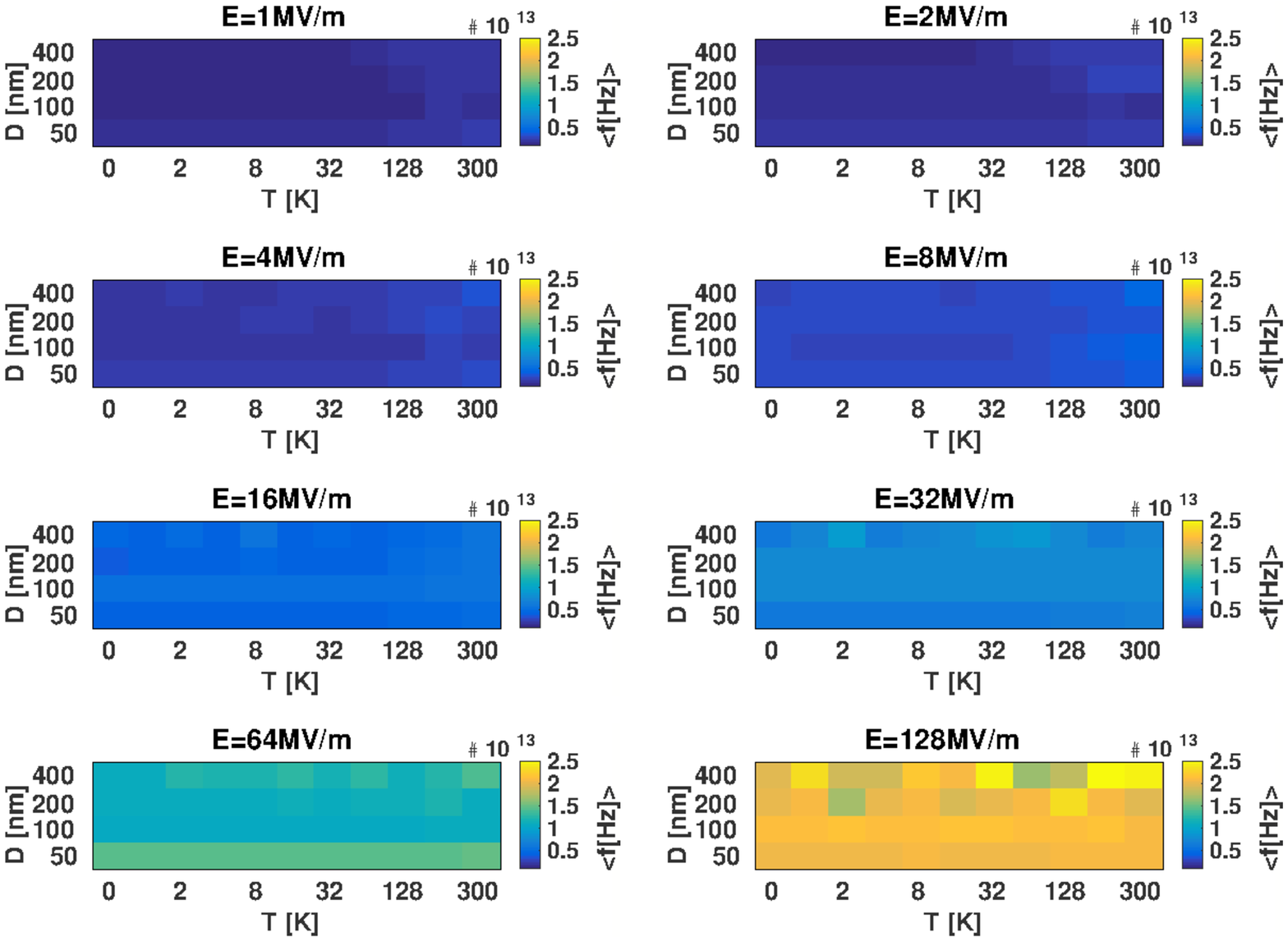}
\end{center}
\vspace{-7mm}
\caption{Average peak frequency $<f[\rm Hz]>$. 
}
\vspace{-6mm}
\label{avg_freq}
\end{figure}

\begin{figure} [h]
\begin{center}
\vspace{-5mm}
\includegraphics [width=8.0 cm] {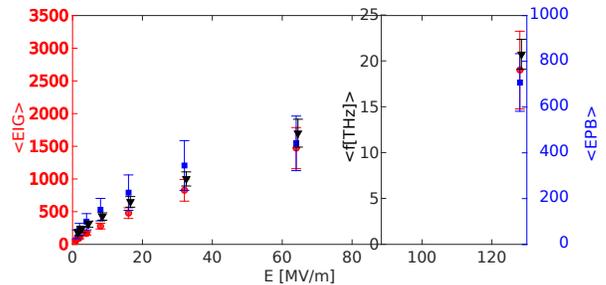}
\end{center}
\vspace{-7mm}
\caption{Average number of electrons in the gap $<\rm EIG>$ (red circles), 
average number of electrons per bunch $<{\rm EPB}>$ (blue squares) 
\red{and averaged frequency $<{\rm f\ [THz]}>$ (black triangles). 
Markers are slightly shifted horizontally for clarity.}
}
\label{eig_epb__VS_E}
\end{figure}

\textit{Frequency.} It has been shown \cite{jonsson2013} that the current frequency
depends only on the electric field for zero emission velocity. That is why all the plots of 
Fig.\ \ref{avg_freq} show
roughly homogeneous color; the frequency doesn't change with changing
temperature, nor gap spacing. 
\red{Apparent variations in peak frequency with increased temperature and gap spacing are numerical artefacts due to the fact that in the parameter regions where the bunches are poorly defined, the exact value of the peak frequency is hard to define.} 
The average number of electrons in the diode gap (EIG) obtained
for all temperatures and gap spacings, but at a fixed electric field,
is shown in Fig. \ref{eig_epb__VS_E}. This number increases simply due
to the increasing space-charge limited current density given by the
classical Child-Langmuir law \cite{langmuir1913}. On the same figure we
can see the average EPB. This was done as a way to cross check our work
with previous work \cite{valfells2002}.

\textit{Conclusion.} THz radiation with microdiodes where several bunches are present in
the gap can be maintained only at temperatures below 10 K. Although
such temperatures are attainable they are impractical for real world
operation.  In this paper we  showed that if the number of bunches
in the system can be  controlled to be close to one, the SNR significantly
improves. Such systems can have a high SNR ($>25$ dB) even
at room temperature. 

\vspace{-3mm}
This work was funded by the Icelandic Research Fund grant 120009021.
\red{AM acknowledges additional support from COST Action MP1204.} We are
thankful to Samuel Perkin from Reykjavik University and Kevin Jensen
from the US Naval Research Laboratory for useful comments.
\vspace{-7mm}


%

\end{document}